\pdfoutput=1
\documentclass[11pt,twoside,a4paper,cmspaper,final,collab]{cms-tdr}
\def\svnVersion{0f443ee}\def\svnDate{2024/10/02}\def\cmsCernNoTag{CERN-EP-2024-143}\def\cmsCernDate{\today}\def\cmsMessage{Published in the Journal of High Energy Physics as \href{http://dx.doi.org/10.1007/JHEP09(2024)186}{\doi{10.1007/JHEP09(2024)186}.}}
\begin{document}\cmsNoteHeader{EXO-21-017}

\newcommand{\Wg}{\ensuremath{\PW\PGg}\xspace}
\newcommand{\Zg}{\ensuremath{\PZ\PGg}\xspace}
\newcommand{\GG}{\ensuremath{\PGg\PGg}\xspace}
\newcommand{\WW}{\ensuremath{\PW\PW}\xspace}
\newcommand{\WWG}{\ensuremath{\PW\PW\PGg}\xspace}
\newcommand{\WZG}{\ensuremath{\PW\PZ\PGg}\xspace}
\newcommand{\tW}{\ensuremath{\PQt\PW}\xspace}
\newcommand{\WZ}{\ensuremath{\PW\PZ}\xspace}
\newcommand{\ptg}{\ensuremath{\pt\left(\PGg\right)}\xspace}
\newcommand{\mx}{\ensuremath{m_{\PX}}\xspace}
\newcommand{\mtsss}{\ensuremath{\mT/\sqrt{s}}\xspace}
\newcommand{\pp}{\ensuremath{\Pp{}\Pp}\xspace}
\newcommand{\cmsParagraph}[1]{\textit{#1: }}
\providecommand{\cmsTable}[1]{\resizebox{\textwidth}{!}{#1}}
\newlength\cmsTabSkip\setlength{\cmsTabSkip}{1ex}

\cmsNoteHeader{EXO-21-017}
\title{Search for a resonance decaying to a W boson and a photon in proton-proton collisions at \texorpdfstring{$\sqrt{s} = 13\TeV$}{sqrt(s) = 13 TeV} using leptonic W boson decays}

\date{\today}

\abstract{A search for a new charged particle \PX with mass between 0.3 and 2.0\TeV decaying to a \PW boson and a photon is presented, using proton-proton collision data at a center-of-mass energy of 13\TeV, collected by the CMS experiment and corresponding to an integrated luminosity of 138\fbinv. Particle \PX has electric charge $\pm1$ and is assumed to have spin 0. The search is performed using the electron and muon decays of the \PW boson. No significant excess above the predicted background is observed. The upper limit at 95\% confidence level on the product of the production cross section of the \PX and its branching fraction to a \PW boson and a photon is found to be 94 (137)\unit{fb} for a 0.3\TeV resonance and 0.75 (0.81)\unit{fb} for a 2.0\TeV resonance, for an \PX width-to-mass ratio of 0.01\% (5\%). This search presents the most stringent constraints to date on the existence of such resonances across the probed mass range. A statistical combination with an earlier study based on the hadronic decay mode of the \PW boson is also performed, and the upper limit at 95\% confidence level for a 2.0\TeV resonance is reduced to 0.50 (0.63)\unit{fb} for an \PX width-to-mass ratio of 0.01\% (5\%).}

\hypersetup{
pdfauthor={CMS Collaboration},
pdftitle={Search for a resonance decaying to a W boson and a photon in proton-proton collisions at sqrt(s)=13TeV using leptonic W boson decays},
pdfsubject={CMS},
pdfkeywords={CMS, wgamma, leptonic, resonances}}

\maketitle 

\section{Introduction}

Many extensions of the standard model (SM) of particle physics, 
such as unified field
theories~\cite{Arkani-Hamed:2016kpz}, two Higgs doublet
models~\cite{Logan:2018wtm}, and folded-supersymmetry
models~\cite{Burdman:2014zta}, predict the existence of charged
resonances decaying to a pair of SM gauge bosons.
In this paper, a search for a particle \PX decaying to a \PW boson and a photon (\Wg) is
presented. Particle \PX has electric charge $\pm1$ and is assumed to have spin 0. The search utilizes proton-proton (\pp) collision data at center-of-mass energy $\sqrt{s}=13~\mathrm{TeV}$ delivered by the CERN LHC and collected with the CMS detector in 2016--2018 with a total integrated luminosity of 138\fbinv. 

Previous searches for \PX decaying to
\Wg (and \Zg, when it is not possible to distinguish
the two massive bosons) have been performed by the ATLAS
Collaboration using \pp collision data at
$\sqrt{s}=8\TeV$~\cite{ATLAS:2014lfk} and
13\TeV~\cite{ATLAS:2018sxj,ATLAS:2023kcu} in leptonic and
hadronic final states.
No signal-like excess of events above background expectations was observed.
Recently, the CMS Collaboration presented a search in the hadronic channel, also using
138\fbinv of 13\TeV data~\cite{CMS:2021zxu}.
An excess of events over background expectations was reported near a
\Wg resonant mass of 1.58\TeV. The local significance was 2.8 standard deviations assuming a narrow resonance ($\Gamma_{\PX}/\mx=0.01\%$, where $\Gamma_{\PX}$ is the width and \mx is the
 mass).  For a broad ($\Gamma_{\PX}/\mx=5\%$) resonance, the local significance was 3.1 standard deviations.
A search for \Wg with leptonic \PW boson decays, which is both
complementary and more sensitive within the range of the reported
excess,  is the focus of this paper.
Additionally, a combination of leptonic and hadronic channels is performed.

\begin{figure}[htp]
\centering
    \includegraphics[width=0.45\textwidth]{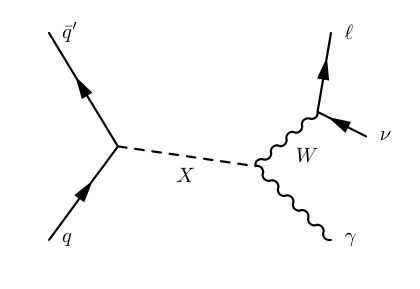}
    \caption{Leading order Feynman diagram for a heavy particle $\PX$ decaying to
      a \PW boson and a photon; the \PW boson subsequentially decays leptonically.
    \label{diagram}}
\end{figure}

Figure~\ref{diagram} depicts the leading order (LO) diagram for the production of \PX with decay to a final state containing a photon $\PGg$, a charged lepton $\Pell$, and a neutrino $\PGn$. 
This search primarily focuses on the \PW boson decaying to $\Pe\PGn$ and $\PGm\PGn$.
The decays of the \PW boson to a tau lepton and a neutrino followed by the decay $\PGt\to\Pell\PGn\PGn$ ($\Pell=\Pe,\PGm$) are also included in the signal sample.
Although the branching fraction of the \PW boson to final states containing $\Pe\PGn$ and $\PGm\PGn$ is small ($\approx$26\%~\cite{ParticleDataGroup:2022pth}), 
this search benefits from a reduced background as compared to the hadronic case, 
especially when studying resonances with masses between 300\GeV and 2\TeV.
The signature of a $\PGn$ is the missing transverse momentum \ptmiss.
Because the longitudinal component of the neutrino momentum cannot be measured or inferred directly, the variable used for the resonance search is the transverse mass \mT, defined using only transverse energies and momenta as follows:
\begin{equation}
    \left(\mT\right)^2 = \left(\ET\left(\PGg\right)+\ET\left(\Pell\right)+\ptmiss\right)^2-\abs{\ptvec(\PGg)+\ptvec(\Pell)+\ptvecmiss}^2,
\end{equation}
where the transverse energy \ET is defined as $\sqrt{\smash[b]{m^2+\pt^2}}$, $\ET\left(\PGg\right)$ is the transverse energy
of the photon, $\ET\left(\Pell\right)$ is the transverse energy of the lepton,  $\ptvec(\PGg)$ is the transverse momentum of the photon, and $\ptvec(\Pell)$ is the transverse momentum of the charged lepton.

The search results are
interpreted in terms of limits on the product of the cross section and the
branching fraction to \Wg within a mass range of 0.3--2.0\TeV
for both narrow and broad resonances. 
The width of the narrow resonance is negligible compared
to the 2--7\% experimental resolution in the considered mass range.
The tabulated results are provided in a HEPData record~\cite{HEPData}.

\section{The CMS detector}
The CMS apparatus~\cite{CMS:2008xjf,CMS:2023gfb} is a multipurpose, nearly hermetic detector, designed to trigger on~\cite{CMS:2016ngn,CMS:2020cmk} and identify electrons, muons, photons, and (charged and neutral) hadrons~\cite{CMS:2020uim,CMS:2018rym,CMS:2014pgm}.
The central feature of the CMS apparatus is a superconducting solenoid
of 6\unit{m} internal diameter, providing a magnetic field of
3.8\unit{T}. Within the solenoid volume are a silicon pixel and strip
tracker, a lead tungstate crystal electromagnetic calorimeter (ECAL),
and a brass and scintillator hadron calorimeter (HCAL), each composed
of a barrel and two endcap sections. Forward calorimeters extend the
pseudorapidity ($\eta$) coverage provided by the barrel and endcap
detectors. Muons are measured in gas-ionization detectors embedded in
the steel flux-return yoke outside the solenoid.

Events of interest are selected using a two-tiered trigger system. The
first level (L1), composed of custom hardware processors, uses
information from the calorimeters and muon detectors to select events
at a rate of around 100\unit{kHz} within a fixed latency of
4\mus~\cite{CMS:2020cmk}. The second level, known as the high-level
trigger (HLT), consists of a farm of processors running a version of
the full event reconstruction software optimized for fast processing,
and reduces the event rate to around 1\unit{kHz} before data
storage~\cite{CMS:2016ngn}.

A more detailed description of the CMS detector, together with a
definition of the coordinate system used and the relevant kinematic
variables, can be found in Ref.~\cite{CMS:2008xjf}.

\section{Data sets and simulated samples}
The data used in this search correspond to a total integrated
luminosity of 138\fbinv at $\sqrt{s}=13\TeV$ and were recorded by the CMS experiment. In years 2016, 2017, and 2018, the data collected
correspond to integrated luminosities of 36, 42, and 60\fbinv,
respectively~\cite{CMS:2021xjt,CMS-PAS-LUM-17-004,CMS-PAS-LUM-18-002}. 

The data selection process relies on a combination of triggers.
Events containing \PW bosons that decay to the $\Pe\PGn$ final state are selected with an isolated single-electron trigger with a transverse momentum \pt requirement of 27 (32)\GeV for 2016 (2017 and 2018).
A single-photon trigger with a \pt requirement of 175 (200)\GeV for 2016 (2017 and 2018) is also used, to recover events where the electron track is not correctly reconstructed by the HLT software at high \pt, causing the electron to be classified as a photon.
The primary trigger for events containing \PW bosons that decay to the
$\PGm\PGn$ final state requires an isolated muon with a minimum
\pt of at least 27 (24)\GeV for the 2017 (2016 and 2018) data.

Simulated signal samples are produced at LO in perturbative quantum chromodynamics (pQCD); the charged particles \PX of the electroweak triplet $\pi_3$, introduced in Ref.~\cite{Arkani-Hamed:2016kpz}, are generated with masses of 300, 350, 400, 450, 500, 600, 700, 800, 900, 1000, 1200, 1400, 1600, 1800, and 2000\GeV,
for both narrow ($\Gamma_X/m_X=0.01\%$) and broad ($\Gamma_X/m_X=5\%$) resonances.
These samples are employed to calculate the signal efficiency and optimize the analysis selection. 
The interference with SM \Wg production is neglected because of its negligible influence on the signal kinematic properties.

The analysis utilizes a background model obtained directly from data.  However, simulated background
events are used in the analysis optimization and to check the robustness of the background
modeling. The dominant background is the
irreducible SM \Wg process. Other nonnegligible background processes are \Zg, other SM multi-bosons (\eg, \GG, \WW, \WZ, \WWG, and \WZG ), $\PV{}+\text{jets}$ where \PV is \PW or \PZ , $\PGg{}+\text{jets}$, \ttbar, and \tW production.
Simulated background processes are produced at next-to-LO (NLO) in pQCD, except for the $\PW{}+\text{jets}$ and $\PGg{}+\text{jets}$ processes, which are produced at LO. 
The cross sections used to normalize the simulated sample to the integrated luminosity of the data for \ttbar and $\PV{}+\text{jets}$ were calculated at next-to-NLO (NNLO) in pQCD, while the other samples use NLO cross sections.

\begin{sloppypar}
The \ttbar and  \tW processes are generated with \POWHEG (v2.0)~\cite{Nason:2004rx,Frixione:2007vw,Alioli:2010xd,Jezo:2015aia}, while all the other signal and background events are simulated with
\MGvATNLO 2.2.2 (2.4.2)~\cite{Alwall:2014hca} for 2016 (2017 and 2018) conditions.
For all simulated samples, the \pp interaction simulation is supplemented with NNPDF3.0
NLO~\cite{NNPDF:2014otw} (NNPDF3.1 NNLO~\cite{NNPDF:2017mvq}) parton
distribution functions (PDFs) for 2016 (2017 and 2018) conditions. 
The event generator \PYTHIA 8.205 (8.230)~\cite{Sjostrand:2014zea} is used
to simulate fragmentation, parton showering, and hadronization, with the CUETP8M1~\cite{Skands:2014pea,CMS:2015wcf} (CP5~\cite{CMS:2019csb})
underlying event tune for 2016 (2017 and 2018) samples.
\end{sloppypar}

All simulated samples are processed with the full CMS detector model based on
\GEANTfour~\cite{EXT:GEANT4-2002}.
The instantaneous luminosity delivered by the LHC results in additional
\pp interactions in the same or neighboring bunch crossings,
known as pileup. The average number of pileup interactions in the 2016
(2017 and 2018) data set is about 23 (32).
Pileup effects are taken into
account by superimposing simulated minimum bias events on the hard
scattering interaction, with a multiplicity distribution reweighted to match that observed in the data~\cite{CMS:2018mlc}.

\section{Event reconstruction}
The particle-flow (PF) algorithm is a global event reconstruction
technique~\cite{CMS:2017yfk} that aims to reconstruct and identify
every particle in an event, with an optimized combination of
information from the various elements of the CMS detector, as a PF candidate. 

Electrons are reconstructed by geometrically matching charged-particle
tracks from the tracking system with energy clusters deposited in the ECAL. 
The reconstruction efficiency is higher than 95\% for $\ET>20\GeV$~\cite{CMS:2020uim}. 
The electron momentum is evaluated by
combining the energy measurement in the ECAL with the momentum
measurement in the tracker. The momentum resolution for
electrons with \pt of ${\approx}45\GeV$ from $\PZ\to\Pe\Pe$
decays is 1.7--4.5\%. It is generally better in the barrel region than
in the endcaps, and it also depends on the bremsstrahlung energy
emitted by the electron as it traverses the material in front of the
ECAL~\cite{CMS:2020uim}.

Muons are reconstructed from compatible tracks in the inner tracker
and the muon detectors~\cite{CMS:2018rym}. They are measured in the
pseudorapidity range $\abs{\eta} < 2.4$ with detection planes made
using three technologies: drift tubes, cathode strip chambers, and
resistive-plate chambers. The single-muon trigger efficiency exceeds
90\% over the full $\eta$ range, and the efficiency to reconstruct and
identify muons is greater than 96\%. 
The momenta of muons are obtained from the curvature of their corresponding tracks.
Matching muons to tracks measured
in the silicon tracker results in a relative \pt
resolution, for muons with \pt up to 100\GeV, of 1\% in the barrel and
3\% in the endcaps. The \pt resolution in the barrel is better than
7\% for muons with \pt up to 1\TeV~\cite{CMS:2018rym}.

The energies of photons are obtained from ECAL measurements.
In the barrel section of the ECAL, an energy resolution of about 1\%
is achieved for unconverted or late-converting photons in the tens of
\GeVns energy range. The energy resolution of the remaining barrel
photons is about 1.3\% up to $\abs{\eta} = 1$, increasing to about 2.5\%
at $\abs{\eta} = 1.4$. In the endcaps, the energy resolution is about
2.5\% for unconverted or late-converting photons and is 3--4\% for all
others~\cite{CMS:2015myp}.

The energies of charged hadrons are determined from a combination of their momenta measured in
the tracker and their corresponding ECAL and HCAL energy deposits,
which are corrected for the response function of the calorimeters to
hadronic showers. Finally, the energies of neutral hadrons are
obtained from the corresponding corrected ECAL and HCAL energies.

Jets are clustered from PF candidates using the anti-\kt
algorithm~\cite{Cacciari:2008gp,Cacciari:2011ma} with a distance parameter of 0.4. Jets found
within $\DR \equiv \sqrt{\smash[b]{(\Delta\eta)^2+(\Delta\phi)^2}}=0.4$ from an identified muon, electron, or photon
are removed, where $\phi$ is the azimuthal angle. Jet flavor tagging is performed using the
\textsc{DeepJet}~\cite{Bols:2020bkb} tagger at the medium working
point. \textsc{DeepJet} is a multi-class flavor tagger algorithm that
combines information from charged and neutral PF
candidates, reconstructed secondary vertices, and global variables
using a deep neural network~\cite{Bols:2020bkb,CMS-DP-2023-005}. The
medium working point is characterized by a 1\% misidentification probability for light-quark or gluon jets and an efficiency of 80\% for tagging jets originating from a $\PQb$ quark. 
Rejecting events with a \PQb jet passing the medium working point reduces the \ttbar backgrounds significantly. 

The missing transverse momentum vector \ptvecmiss is computed as the negative vector sum of the transverse momenta of all the PF candidates
in an event, and its magnitude is denoted as
\ptmiss~\cite{CMS:2019ctu}. The \ptvecmiss is modified to account for
corrections to the energy scale of the reconstructed jets in the event.

The primary vertex (PV) is taken to be the vertex corresponding to the hardest scattering in the event, evaluated using tracking information alone, as described in Section 9.4.1 of Ref.~\cite{CMS-TDR-15-02}.

\section{Event selection}

Events are selected in two steps. A basic selection is applied with loose lepton and photon selection criteria to veto backgrounds with multiple leptons and photons. Then a tighter event selection is applied to improve the signal sensitivity.

In the basic selection, candidate electrons, muons, and photons are selected using the standard CMS ``medium''
photon and electron and ``tight'' muon identification (ID) requirements~\cite{CMS:2020uim,CMS:2018rym}. Many of the kinematic variable thresholds are the same for both lepton
channels because their signal kinematic distributions  and  background compositions are very similar. 

Leptons are required to have $\pt>10\GeV$, and
photons $\pt>15\GeV$. Leptons are also required to satisfy $\eta$ selections to ensure they are in the regions of the detector where they will be well measured.
Since the leptons from the signal process are typically isolated, backgrounds from jets being misidentified as an electron are reduced by requirements on $\mathcal{I}_{\mathrm{PF,rel}}$, defined as the summed \pt of reconstructed PF candidates within a cone centered on the electron momentum vector divided by the electron \pt.
Similar isolation criteria are integrated into the CMS standard muon and photon identification requirements.
Leptons and
photons must be separated by $\DR > 0.4$.
These kinematic and identification requirements are referred to
as the ``basic object selections'' and are summarized in Table~\ref{tab:objsel}. Leptons and photons passing these requirements are referred to as basic electrons, muons, and photons later.

To reduce backgrounds from \ttbar and multi-boson processes, events with more than one lepton or photon passing the basic requirements are rejected. 

\begin{table}[ht]
  \centering
  \topcaption{Basic object selection requirements. Definitions are described in more detail in the text.}
\label{tab:objsel}
  \cmsTable{
\begin{tabular}{lccc}
Selection                & Electron                              & Muon                   & Photon       \\
\hline
\pt                      & $>$10\GeV                           & $>$10\GeV            & $>$15\GeV  \\
$\eta$                   & $\abs{\eta}<1.44$ or $1.57<\abs{\eta}<2.1$ & $\abs{\eta}<2.4$  & $\abs{\eta}<1.44$ or $1.57<\abs{\eta}<2.5$\\
ID                       & Medium                                & Tight                  & Medium \\
Isolation                & Integrated with ID                    & $\mathcal{I}_{\mathrm{PF,rel}}<0.15$    & Integrated with ID \\
\DR                      & $\DR(\PGg,\Pe)>0.4$                   & $\DR(\PGg,\PGm)>0.4$   & $\DR(\PGg,\Pe)>0.4$ and $\DR(\PGg,\PGm)>0.4$ \\
\end{tabular}
}
\end{table}

To further reduce backgrounds, an optimization is performed
on additional kinematic variables. 
The selected electron and muon are required to match the trigger object that caused the event to be selected by the HLT.
The lepton \pt requirements are increased to 35 (30)\GeV in the electron (muon) channels to 
exclude events near the trigger threshold where the trigger efficiency changes significantly.
The CMS standard ``tight'' identification requirement on the electron is used~\cite{CMS:2020uim}. 
 A threshold of 40\GeV on \ptmiss is required in both
channels.

For the signal, there is an approximately linear
relationship between \ptg and \mT. This feature can be used to select signal-like events and reject background-like ones.
Requiring $\ptg>0.4~\mT$ rejects SM \Wg process events, which
generally tend to have a softer \ptg spectrum, while requiring
$\ptg<0.55~\mT$ preferentially rejects  \ttbar events. 
These \mT-dependent \ptg selection cuts achieve a background rejection of more than 55\% while preserving about 90\% of signal events.
The requirement on the photon pseudorapidity is also tightened to $\abs{\eta}<1.44$,
thus selecting only photons reconstructed in the barrel section of the calorimeters.
To further
reject \ttbar backgrounds, events containing $\PQb$-tagged jets passing the
medium working point of the \textsc{DeepJet} tagger are rejected.

In the electron channel, to suppress $\PZ \to \Pe\Pe$
events where one of the electrons is mistakenly reconstructed as a
photon, a pixel detector seed veto (PSV), which rejects photon candidates with more than two hits in the pixel detector, is applied~\cite{CMS:2020uim}. 
The contribution of this
process is further reduced by rejecting events where the invariant
mass of the electron and photon is near the \PZ boson
mass---specifically between 71 and 111\GeV, corresponding to roughly $\pm5$ widths of a reconstructed \PZ boson peak.
These selections are summarized in Table~\ref{tab:evtsel}.

\begin{table}[ht]
\centering
\topcaption{Event selection requirements for the electron and the muon
  channels. Definitions are described in more detail in the text.}
\label{tab:evtsel}
\begin{tabular}{lcc}
Selection           & Electron channel                  & Muon channel       \\
\hline
\Pe/\PGm sel.          & $\pt(\Pe)>35\GeV$ and tight ID                      & $\pt(\PGm)>30\GeV$  and tight ID  \\
\ptmiss             & \multicolumn{2}{c}{ $>$40\GeV }        \\
\PGg sel.          & \multicolumn{2}{c}{ $0.4~\mT<\ptg<0.55~\mT$ and $\abs{\eta}<1.44$  }  \\
Z veto & $\abs{m_{\Pe\PGg}-91.0}>20\GeV $ and PSV     & \NA    \\
\PQb veto       & \multicolumn{2}{c}{No medium \PQb-tagged jets} \\
\end{tabular}
\end{table}

\begin{figure}[htp]
\centering
    \includegraphics[width=0.49\textwidth]{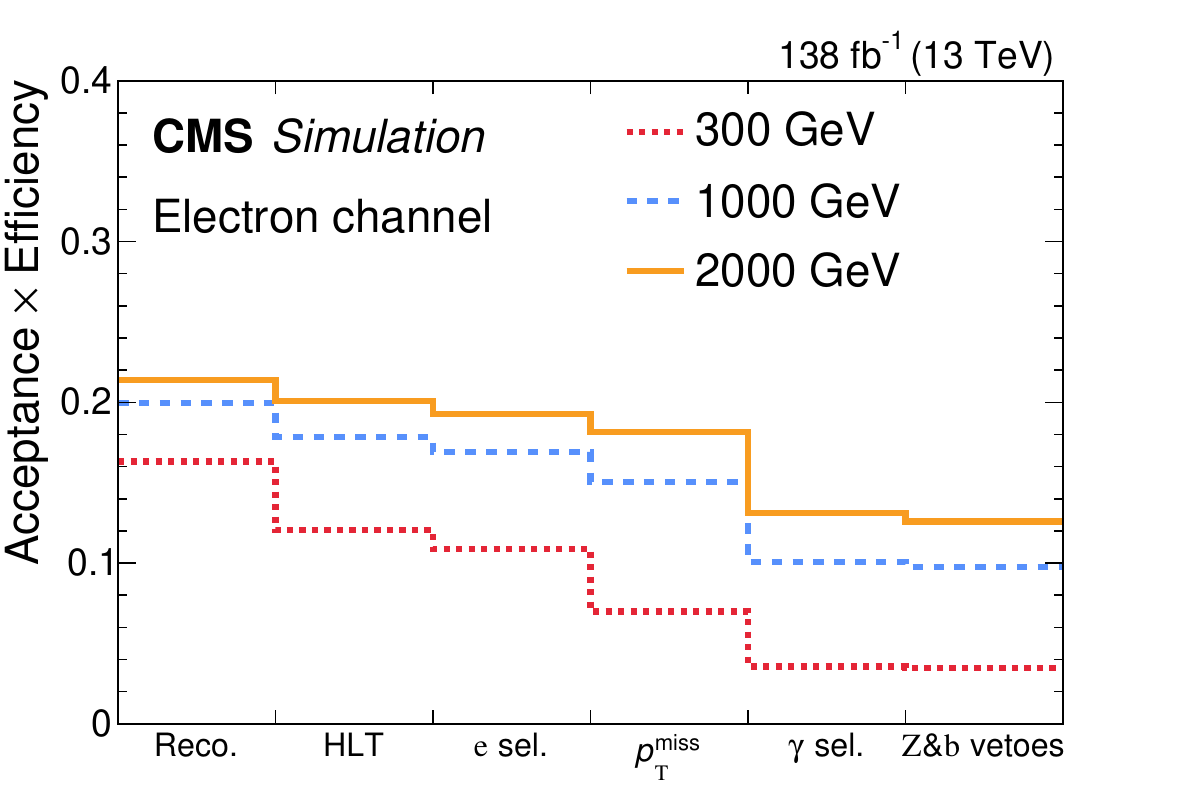}
    \includegraphics[width=0.49\textwidth]{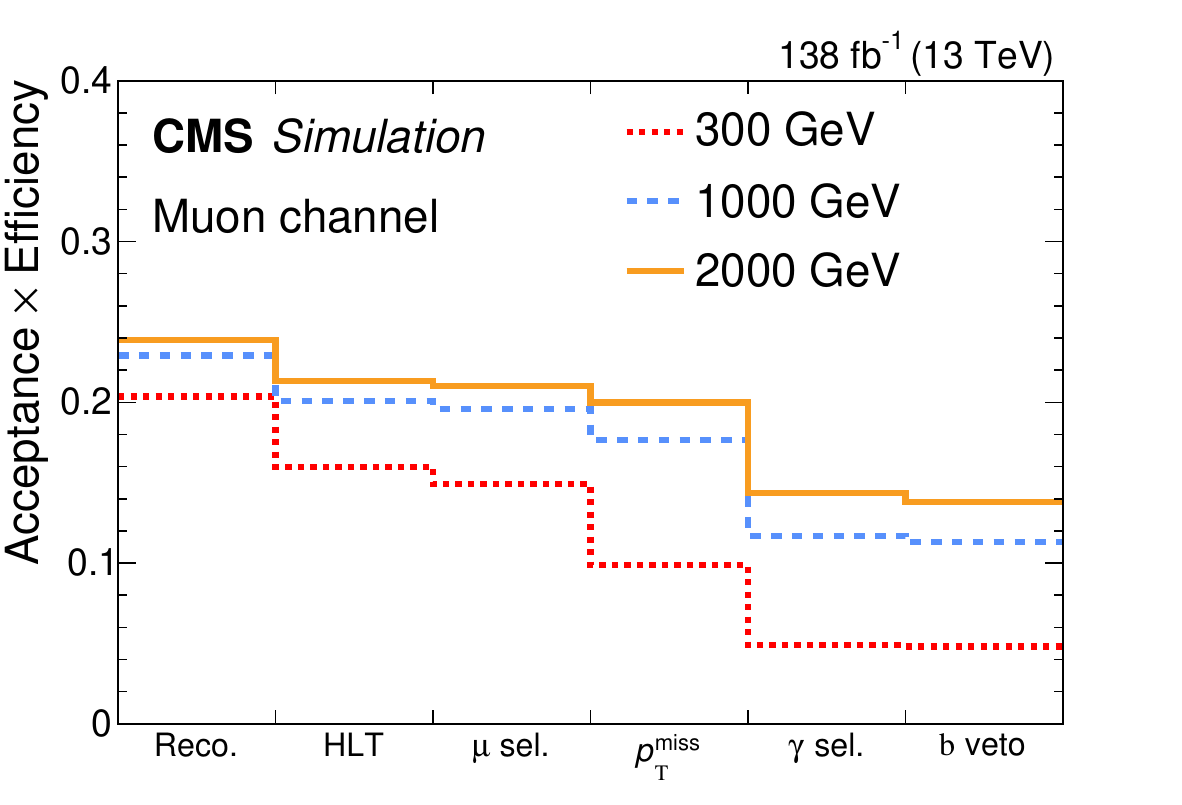}
    \caption{Product of detector acceptance and analysis selections efficiency for different particle mass assumptions---300, 1000, and 2000\GeV, in red, blue, and orange, respectively---to pass sequential requirements in the broad-width case. The narrow-width case is similar. The first bin represents selecting events with exactly one reconstructed basic electron or muon and one photon, satisfying the selection criteria reported in Table~\ref{tab:objsel}. The next bin contains events satisfying the HLT, and the subsequent bins have the selections listed in Table~\ref{tab:evtsel} applied sequentially. The left (right) plot is for the electron (muon) channel.}
    \label{fig:bkgcuttable}
\end{figure}

Signal efficiencies estimated from the simulation of leptonically decaying \PW bosons are shown in Fig.~\ref{fig:bkgcuttable}.  
Events in the first bin require exactly one reconstructed basic electron or muon and one photon, satisfying the selection criteria reported in Table~\ref{tab:objsel}. The next bin contains events satisfying the HLT, and the subsequent bins have the selections listed in Table~\ref{tab:evtsel} applied sequentially.
The tight photon selections, which are necessary to reduce the backgrounds, have the largest impact on the efficiency. 
The narrow (0.01\%) and broad (5\%) resonances exhibit similar
efficiencies within the displayed mass range. 
Events containing \PW bosons that decay to the $\PGt\PGn$ final state with $\PGt$ subsequently decaying to $\Pe$ or $\PGm$ are considered as signal and included in the acceptance. They constitute about 10\% of the selected events.

\begin{figure}[htp]
\centering
    \includegraphics[width=0.49\textwidth]{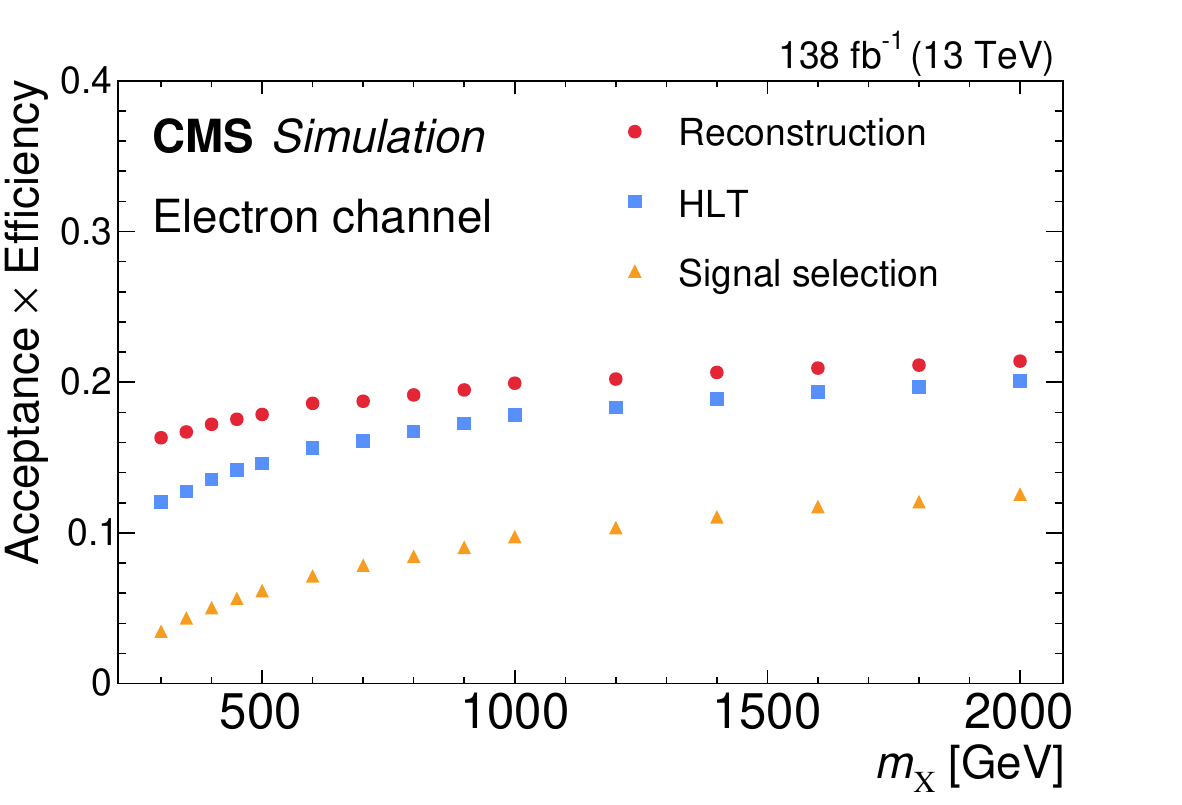}
    \includegraphics[width=0.49\textwidth]{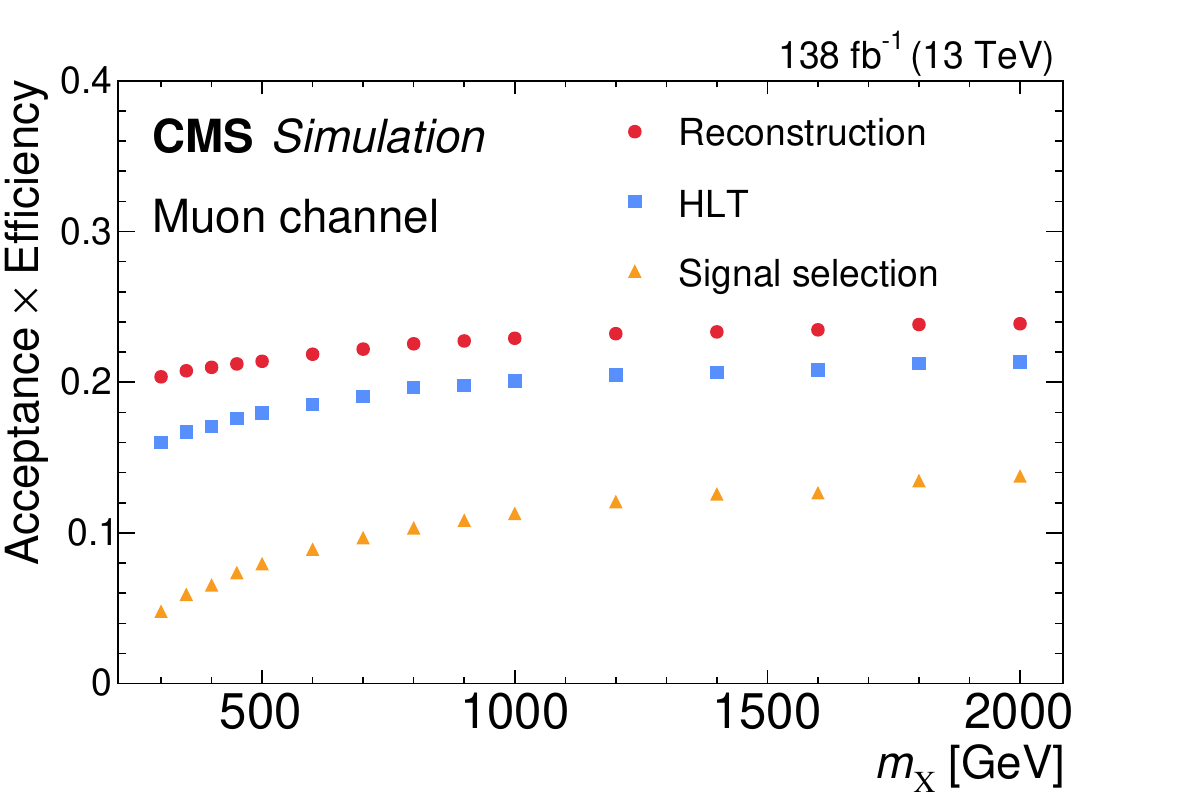}
    \caption{Product of detector acceptance and analysis selection efficiency in the electron (left) and muon
      (right) channel as functions of the particle \PX mass in the broad-width case. The narrow-width case is similar.  Three analysis requirements are applied consecutively:
      event reconstruction, HLT, and final signal selection. The
      product of detector acceptance and analysis selection efficiencies are shown at each stage in red, blue, and orange,
      respectively.
    \label{fig:eff_signal}}
\end{figure}

Signal efficiencies for all simulated signal samples,  separately for the electron and muon
channels, are shown in
Fig.~\ref{fig:eff_signal}. The product of detector acceptances and analysis selection efficiencies are
shown at three consecutive stages: event reconstruction, HLT,
and final signal selection. The total signal efficiency ranges from
3.4 (4.8)\% to 12.7 (13.9)\% for the electron (muon) channel. The
efficiency increases with \mx. The efficiencies
for the electron channel are generally 1--2\% lower than those for the muon channel
owing to differences in the selection criteria applied at each stage of the event selection, as listed in Tables~\ref{tab:objsel} and ~\ref{tab:evtsel}. 
The pseudorapidity and \pt requirements are stricter for electrons than for muons, though this effect becomes less significant at higher resonant masses where lepton distributions are more central. Additionally, the efficiency of tight ID selection for electrons is lower than that for muons~\cite{CMS:2020uim,CMS:2018rym}. The trigger efficiency is also lower for the electron channel at low resonant mass but increases at high resonant mass because of the inclusion of high-\pt single-photon triggers.

\begin{figure}[htp]
\centering
    \includegraphics[width=0.49\textwidth]{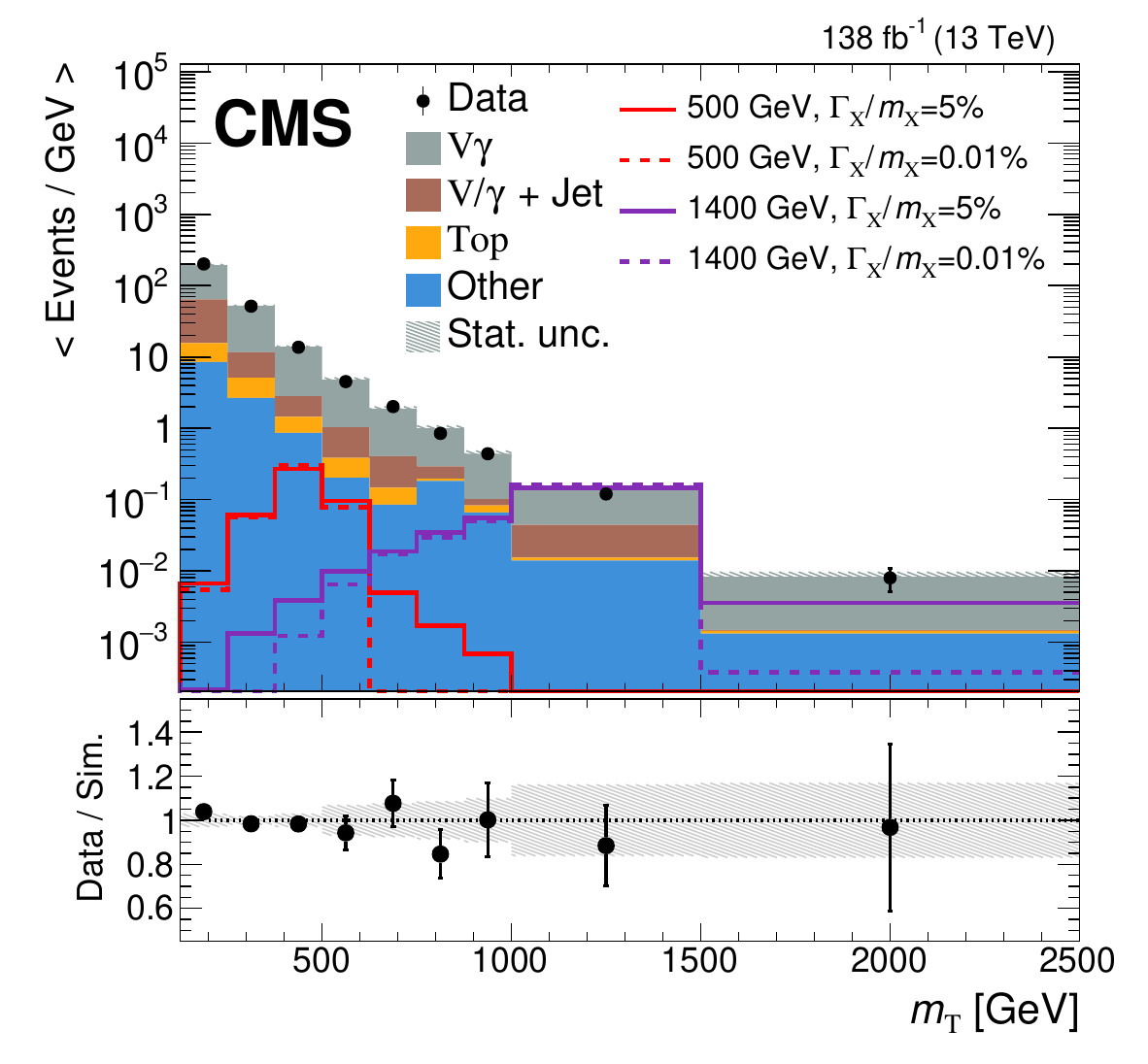}
    \includegraphics[width=0.49\textwidth]{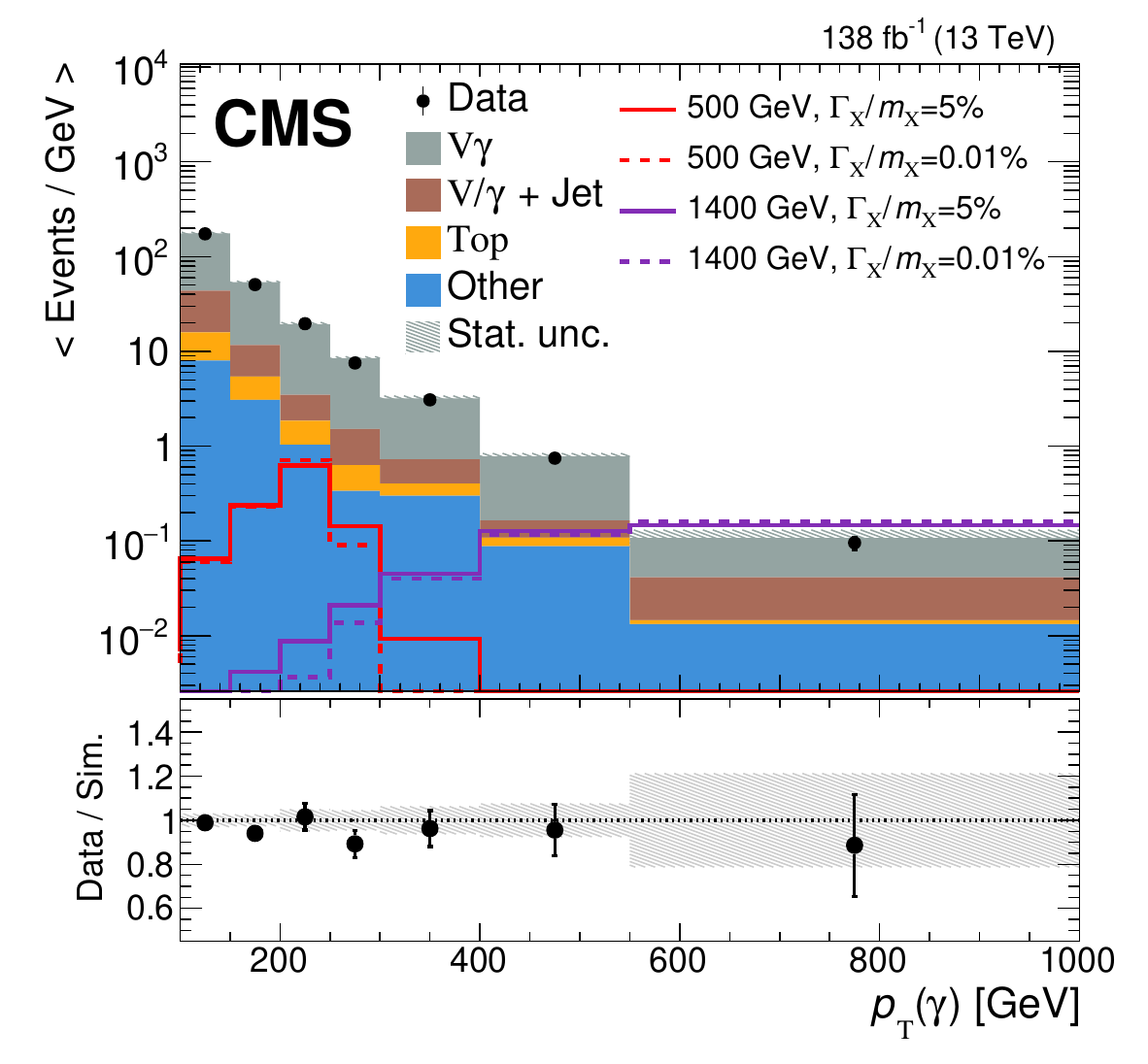}
    \caption{Distributions of \mT (left) and \ptg (right) from
      simulation (stacked histograms) and data events
      (black points) passing all analysis selections. The number of
      events is divided by the width of each individual bin. The
      simulation distributions agree with data within statistical
      uncertainty. Four signal distributions, with two mass assumptions
      and two width assumptions, are also overlaid  (dashed and solid
      lines). Each signal is plotted with a total cross section of
      3\unit{fb}. Because of the limited number of simulated events, in the high
      mass region, progressively larger bin sizes are
      used. The last bin includes the overflow events.
      The lower panel shows the ratio of data to simulation; the hatched band represents the statistical uncertainty in the total background.
    \label{fig:mc_vali}}
\end{figure}

The merged distributions from the electron and muon channels for \mT and \ptg are shown in Fig.~\ref{fig:mc_vali}. 
The background processes are difficult to model in the region where the detection efficiency is increasing because of threshold effects. To avoid this problem, \mT is required to be greater than 230\GeV.
The observed data are consistent with the background expectation from simulation,
and the kinematic distributions in both channels are similar.
The dominant background is the SM $\PV\PGg$ process (where \PV is \PW or \PZ), followed by $\PV/\PGg{}+\text{jets}$ ($\PV{}+\text{jets}$, $\PGg{}+\text{jets}$), and top (\ttbar, \tW) backgrounds. Other background sources contributing to the distributions are grouped under the label ``Other''. 

\section{Signal and background modeling}
\label{sec:modeling}

The \mT distributions from simulations corresponding to the three data-taking periods (\ie, 2016, 2017, and 2018) are similar, with only minor variations reflecting the differences in the 
modeling of the pileup, the trigger requirements, and the detector performance. 
To reduce the signal modeling uncertainties, the simulated signal samples from the three data-taking periods are combined. 
The resulting distribution is fit with a double-sided Crystal Ball (DSCB)
function~\cite{CrystalBallRef,CrystalBallRef2}.
This six-parameter function has a central core with a Gaussian distribution characterized by two parameters: 
 mean transverse mass
$\overline{\mT}$ and resolution $\sigma_{\mT}$.
The \mT distribution obtained from this combined fit is consistent with that obtained from each of the three individual data-taking periods.
The best-fit parameters are then extracted and modeled as a function of the resonant mass, allowing signal templates to be generated at 50\GeV intervals for resonances with masses between 300\GeV and 2000\GeV.

The background simulation samples shown in Fig.~\ref{fig:mc_vali} play a crucial role in our selection optimization and
provide insights into the underlying processes.
However, the background estimation does not rely on the simulation.  
Instead, it is modeled analytically via a fit to the data 
using functions described below:
\begin{equation}
\label{func:dijet}
    p_{0}x^{\sum_{i=1}^{N}p_{i}\log^{i-1}(x)},
\end{equation}
\begin{equation}
\label{func:f1}
    p_{0}\frac{(1-x)^{p_{1}}}{x^{\sum_{i=2}^{N} p_{i}\log^{i-2}(x)}},
\end{equation}
\begin{equation}
\label{func:f2}
    p_{0}\re^{p_{1}x}x^{\sum_{i=2}^{N} p_{i}\log^{i-2}(x)},
\end{equation}
where $x$ is \mtsss. The parameters $p_{i}$ vary freely in the fit, and the index $i$ ranges up to the order $N$ of each function. 
These functions are commonly used in similar searches~\cite{ATLAS:2014lfk,ATLAS:2018sxj,ATLAS:2023kcu,CMS:2021zxu}.
The number of free parameters in each functional form is determined using a Fisher F-test~\cite{ftest}.
The functional form is determined based on data, separately for each data-taking year and lepton channel, and the number of parameters determined by this procedure is either three or four. 
To increase the flexibility of the background modeling,
a discrete profiling method~\cite{Dauncey_2015} is employed to consider all functions when
performing the ultimate fit to the data.
This approach optimizes the background parameters of each function independently to determine which function represents the background best.
 
\begin{figure}[htp]
\centering
    \includegraphics[width=0.49\textwidth]{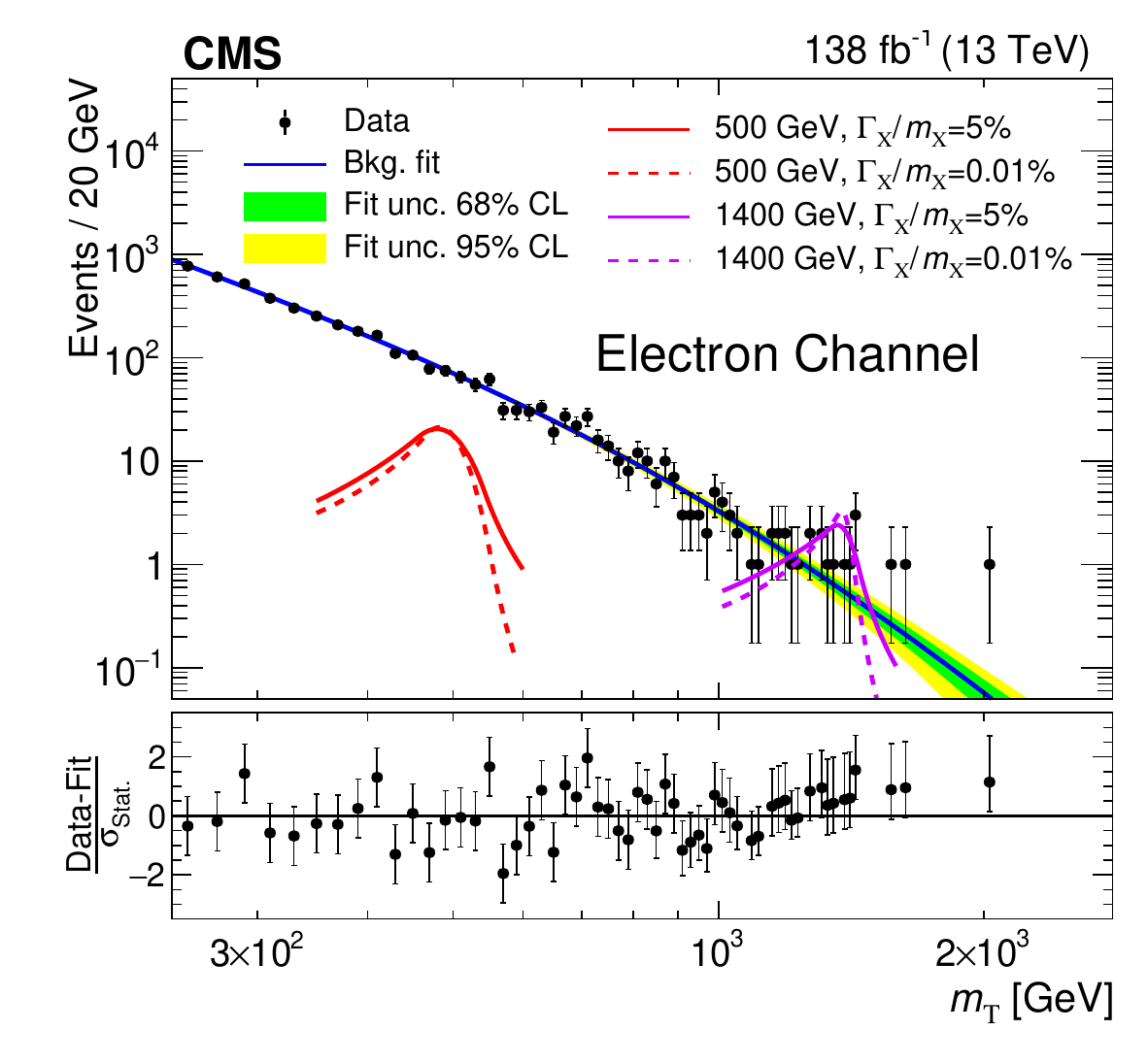}
    \includegraphics[width=0.49\textwidth]{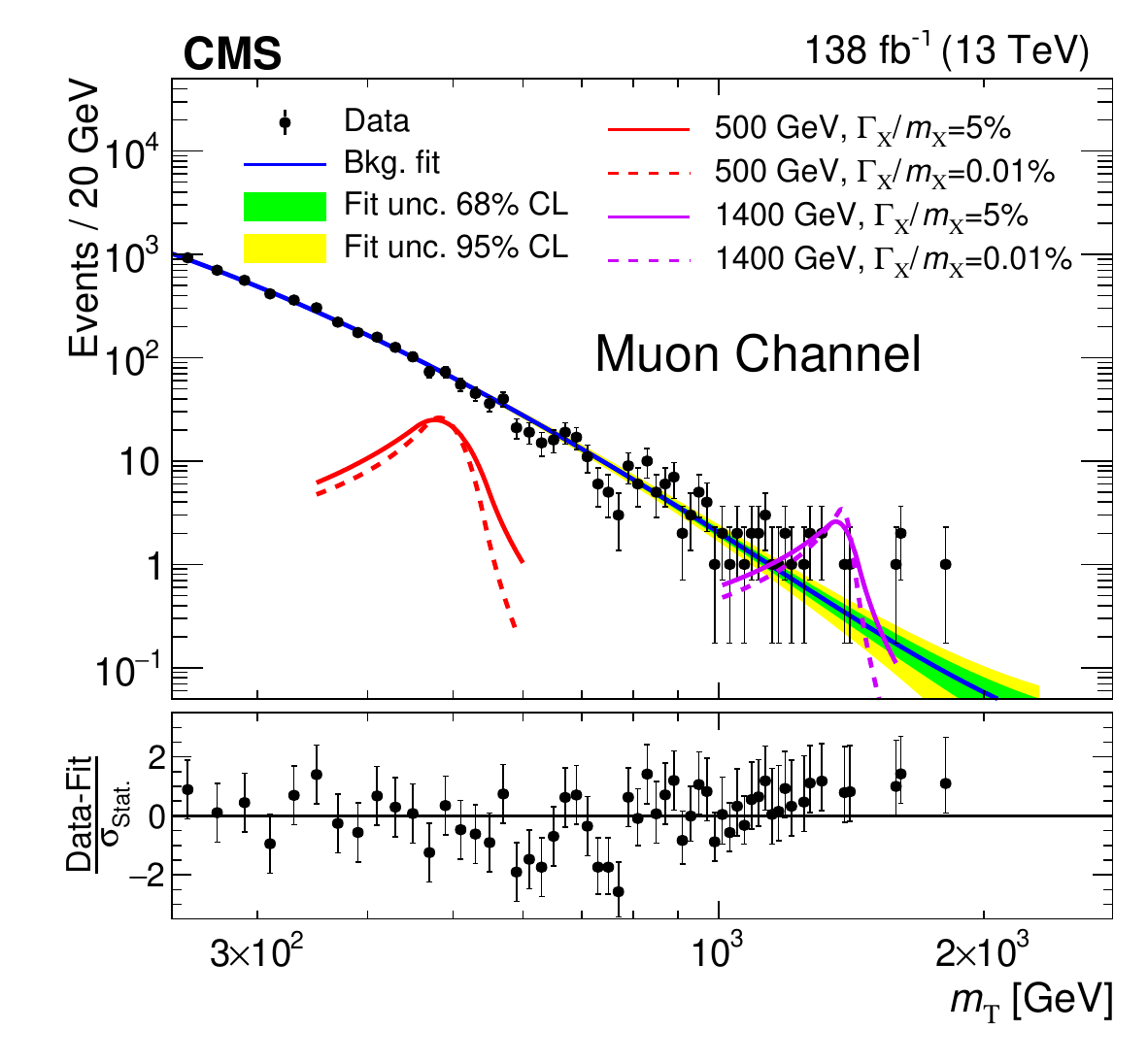}\\
    \caption{\label{fig:bkg_fit} Background-only fit to data (black
      points) with the fitted background model shown as a blue line.
      The green (inner) and yellow (outer) bands show, respectively,
      the 68\% and 95\% confidence level statistical uncertainties in the fitted
      model. Four signal models are also overlaid. Their total cross sections are set to the expected limits at 95\% confidence level from this search with leptonic decays of the W boson.
      There are two mass assumptions---500 and 1400\GeV, in red and magenta,
      respectively---and two width assumptions---narrow (solid curves)
      and broad (dashed curves). The lower panel contains the pull
      distribution, defined as the difference between the data yield
      and the background prediction divided by their combined
      uncertainty. The left (right) panel is the electron (muon)
      channel with all three years' data combined.}
\end{figure}

The fit to the data with the background-only hypothesis is shown in
Fig.~\ref{fig:bkg_fit} for both the electron and the muon channels, with the bin size of 20\GeV.
Several values of bin size from 10 to 30\GeV were tested, and no differences were found in the upper limits.
The fit parameters have no correlation among the years and all are set
to be freely floating. This smooth background parametrization
provides an adequate description of the data over the entire mass range
probed.
 
\section{Systematic uncertainties}

Systematic uncertainties affect both the normalization and shape of
the signal distributions. They are included in the fit in the form of
log-normal (normalization uncertainty) or Gaussian (shape uncertainty) priors for the nuisance parameters. 

The primary systematic uncertainties in the signal normalization are uncertainties
on the integrated luminosity, the PDFs, the choices of
renormalization and factorization scales, the lepton trigger and
identification efficiencies, the photon
identification efficiency, the pileup description, the \PQb tagging efficiency, and an ECAL trigger mistiming correction~\cite{CMS:2020cmk}. 
These systematic uncertainties increase for higher resonance masses and are independent of the width-to-mass ratio.

\cmsParagraph{Lepton and photon reconstruction and identification}
  uncertainties arising from lepton reconstruction and identification
  criteria, as well as trigger efficiencies, are included as
  normalization uncertainties. Muon-related uncertainties range from
  0.5 to 1.2\%~\cite{CMS:2018rym}, while electron-related uncertainties are higher,
  spanning from 1.5 to 3.8\%~\cite{CMS:2020uim}. The photon identification uncertainty is 
  significant in both the electron and muon channels. It is included as a normalization uncertainty, ranging from 1.4 to 5.2\%~\cite{CMS:2020uim}.

\cmsParagraph{Integrated luminosity} the integrated luminosities for the
  2016, 2017, and 2018 data-taking years have 1.2--2.5\% individual
  uncertainties~\cite{CMS:2021xjt,CMS-PAS-LUM-17-004,CMS-PAS-LUM-18-002},
  while the overall uncertainty for the 2016--2018 period is 1.6\%. These are included as normalization uncertainties.

\cmsParagraph{Renormalization/factorization scales and PDFs} theoretical
  uncertainties arise from the choice of the renormalization and
  factorization scales in the matrix elements calculation and the
  proton PDFs. These uncertainties are estimated by varying
  the renormalization and factorization scales independently by factors of 0.5 and 2, excluding the extreme cases (0.5,2) and (2,0.5), and
  assessing the uncertainty to be the largest difference from the
  nominal distribution of \mT of the signal. Additionally, signal
  distributions are generated using different sets of
  NNPDF3.1~\cite{NNPDF:2014otw,NNPDF:2017mvq} replicas to account
  for the PDF uncertainties. The combined uncertainty from these sources
  varies from 1.9 to 4.4\% as normalization uncertainties.

\cmsParagraph{Minor uncertainties} systematic uncertainty from pileup
  reweighting is quantified by varying the total inelastic cross section
  within its $\pm 5\%$ uncertainty range~\cite{CMS:2018mlc}. 
  The data-to-simulation ratios of the \PQb jet tagging efficiency and misidentification probability are used to apply event weights to simulated samples. The uncertainties on these ratios result in uncertainties on the weights, from which systematic uncertainties are inferred.
  These sources, as well as the ECAL trigger mistiming correction
  uncertainty~\cite{CMS:2020cmk} and jet energy scale and resolution uncertainties,
  contribute less than 1\% to the signal normalization uncertainty.

\begin{figure}[htp]
\centering
    \includegraphics[width=0.7\textwidth]{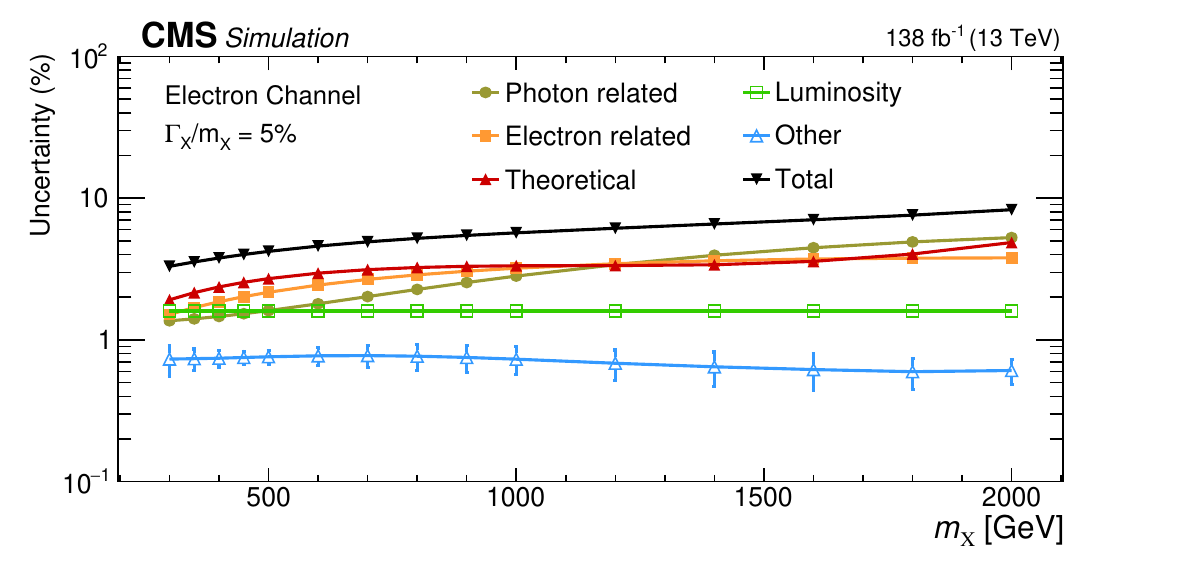}
    \includegraphics[width=0.7\textwidth]{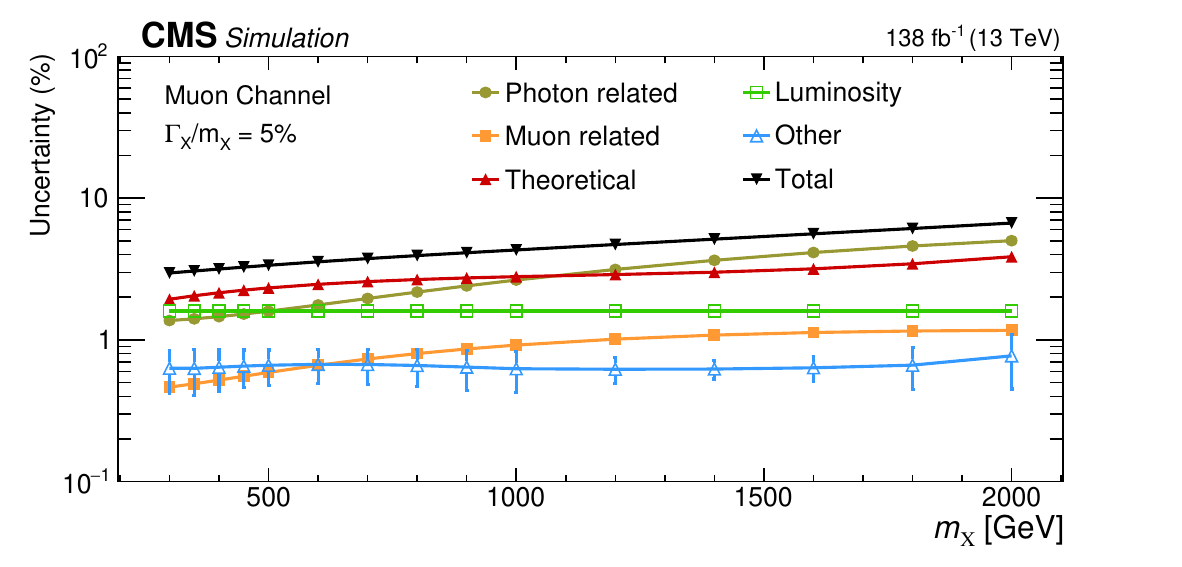}\\
    \caption{\label{fig:syst} Systematic uncertainties affecting the
      signal normalization in the electron (upper) and muon (lower)
      channels for the broad resonance are shown. Error bars represent the asymmetric uncertainties (downward and upward variations), with the central value representing the average uncertainty.
      The total systematic uncertainty (black) increases steadily with the resonant mass assumption. The
      uncertainties for the narrow-width case are nearly identical.}
\end{figure}

A summary of the systematic uncertainties affecting the signal normalization is shown in Fig.~\ref{fig:syst}
for each lepton channel separately.  Uncertainties are
divided into five categories: photon-related, lepton-related, theoretical, integrated luminosity, and all other uncertainties (shown as ``Other''). Their quadrature summation is also included (``Total''). Driven by the increased photon-related uncertainty, the relative signal yield uncertainty grows from 3.2 (2.9) to 8.3 (6.7)\% for resonance masses from 300\GeV to 2\TeV, in the electron (muon) channel.
The narrow and broad resonances have similar values for all uncertainties. 

The signal distribution, modeled by the DSCB function, is primarily
characterized by two important parameters from the Gaussian distribution part: the mean transverse mass
$\overline{\mT}$ and resolution $\sigma_{\mT}$. 
Uncertainties in lepton, photon, and jet energy scales and resolutions can affect the shape of the
signal distribution. Their effect is included in the fit as an additional uncertainty in $\overline{\mT}$ and $\sigma_{\mT}$ that varies as a function of the resonance mass.

To assess the effects of systematic uncertainties in these parameters,
alternative signal \mT distributions are generated by varying the
fundamental kinematic quantities, \eg, photon \pt, lepton \pt, \etc, within their systematic
uncertainty and reevaluating \mT for each event. The difference in the resulting
signal fit parameters $\overline{\mT}$ and $\sigma_{\mT}$ are taken as one-standard-deviation uncertainties. All of these
uncertainties are combined in quadrature, and their combined effect is much smaller than the normalization uncertainties. In most cases, the total
uncertainty is predominantly driven by the variation in the photon
energy scale. The uncertainty in $\overline{\mT}$ is $0.3\%$ at the lowest considered mass and 1.9\% at the highest, while the uncertainty in
$\sigma_{\mT}$ varies from 3 to 7 (3 to 4)\% for narrow (broad)
resonance signals.

As previously described in Section~\ref{sec:modeling}, we treat the
normalization and shape of the background distribution as free
parameters and evaluate them during the fit to the data. 
These parameters are individually evaluated for each signal mass and width hypothesis.

A dedicated study focused on systematic effects affecting the background shape is conducted to confirm that no systematic bias on the signal extraction is present. Pseudo-data are generated from the background model
to produce many \mT spectra---some containing signal and
others without. Each of these spectra is fit in the same fashion as
the data using the nominal discrete-profiling background model. During
these fits, both the signal strengths and the background normalizations
are allowed to vary freely. All distributions of fitted signal
strength minus injected signal strength divided by the signal strength
uncertainty---typically called a ``pull'' distribution---are centered
around zero and have width of unity. Therefore, any potential systematic bias arising from our
background fitting procedure is negligible compared to the statistical
uncertainties inherent in the fit.

In this analysis, only the photon reconstruction and identification uncertainties, integrated luminosity uncertainty, theoretical uncertainties, and the minor uncertainties that are itemized above are correlated between the electron and muon channels. The electron and muon reconstruction and identification uncertainties, and the signal and background model uncertainties are uncorrelated between the electron and muon channels. 
Most uncertainties are correlated across the data-taking years, as the $\mT$ distributions are similar, except for the integrated luminosity uncertainty (which is only partially correlated), the ECAL trigger mistiming correction uncertainty, and the parameters of the background model.
This analysis is dominated by statistical uncertainties; the overall impact of the systematic uncertainties is small.

\section{Results}
\label{Results}

No signal-like excess of events above background expectations is observed. 
Model-specific upper limits are set on the product of the \PX production
cross section and its
branching fraction to \Wg 
for both the narrow and broad resonance scenarios. We determine these
limits with the  \CLs
method~\cite{CLS2,CLS1,CMS-NOTE-2011-005} and the asymptotic
approximation~\cite{Cowan:2010js} using the CMS statistical analysis tool \textsc{Combine}~\cite{CAT-23-001}.
The test statistic is derived from a
likelihood ratio within the asymptotic approximation.

\begin{figure}[htp]
  \centering
  \includegraphics[width=0.49\textwidth]{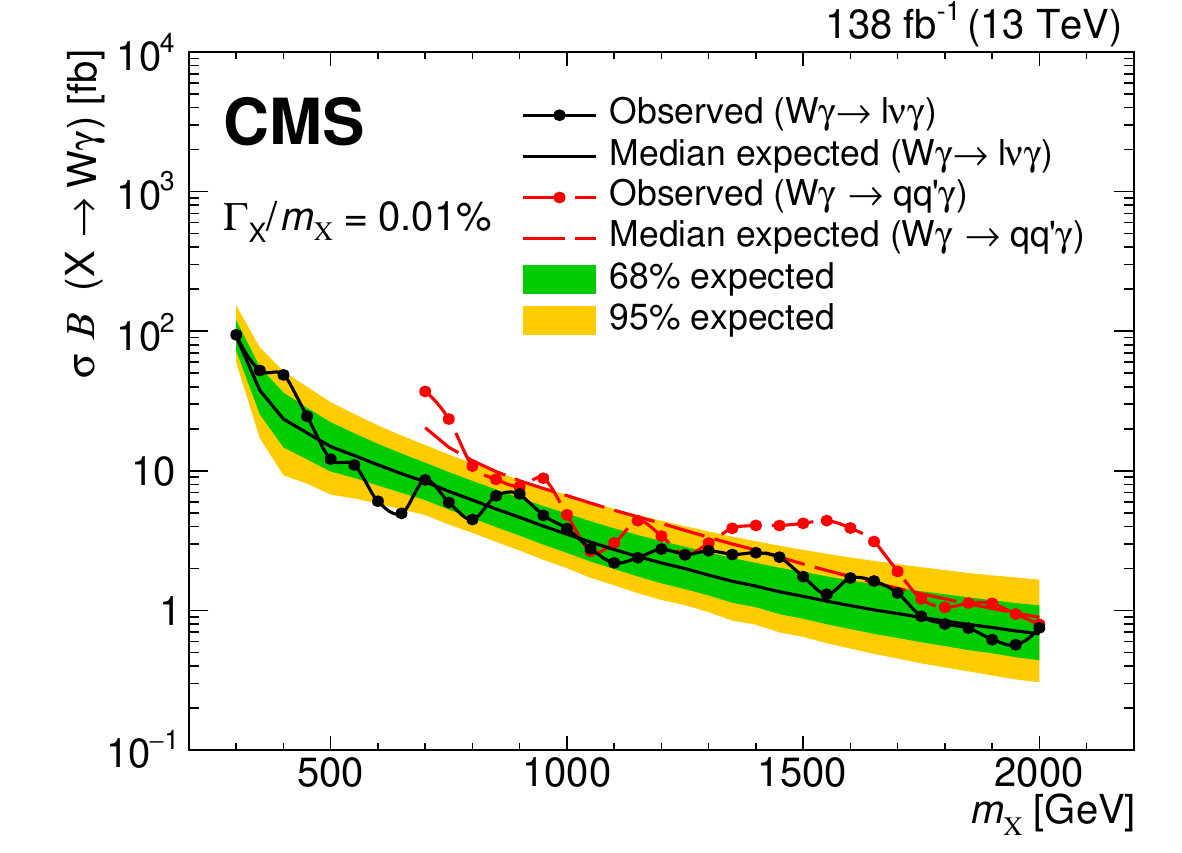}
  \includegraphics[width=0.49\textwidth]{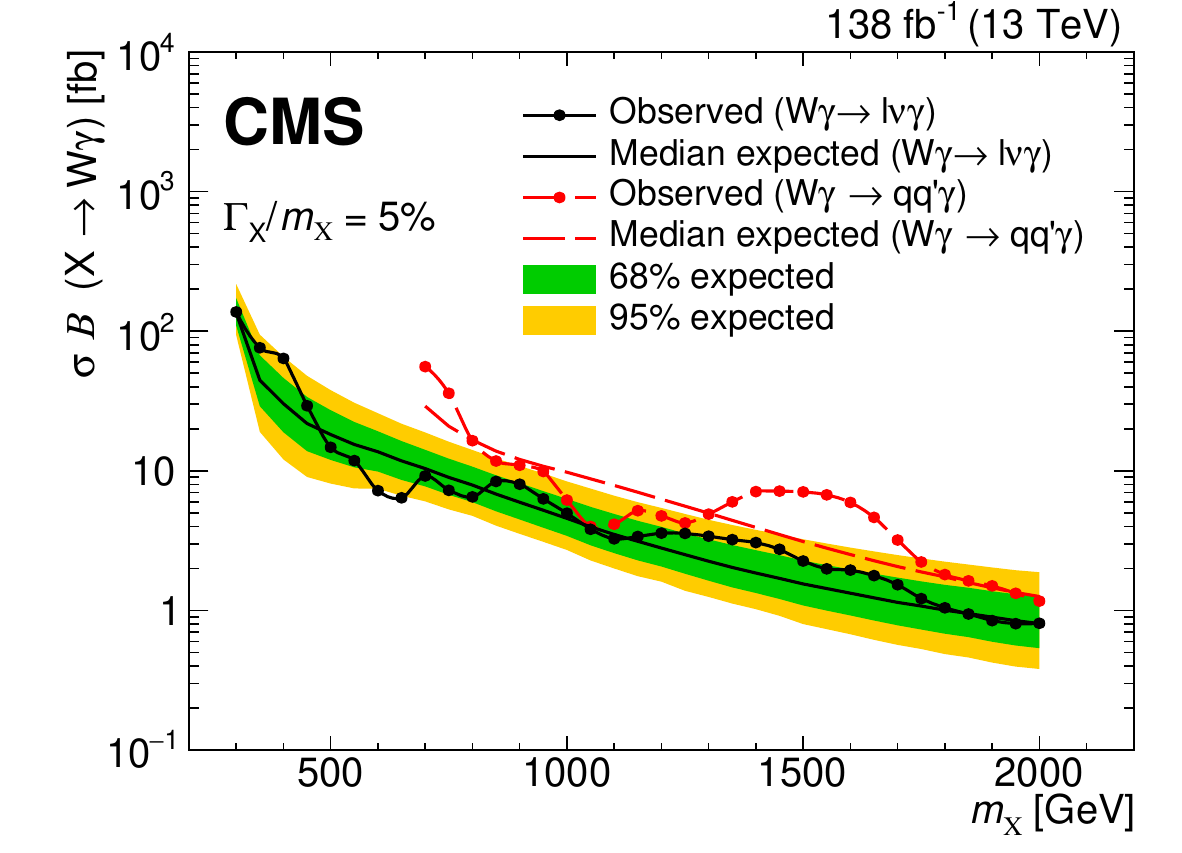}
  \caption{\label{fig:Asymptotic_expected_limits} Expected and observed limits at 95\% \CL on $\sigma\mathcal{B}(X\to\Wg)$ from events with leptonic decays (solid black lines) of the \PW boson as a function of the \PX resonant mass. The limits from the hadronic decays (dashed red lines) of the \PW boson are taken from Ref.~\cite{CMS:2021zxu} and included for comparison.
    The results for the narrow (broad) width assumption are shown on the left (right).}
\end{figure}

Figure~\ref{fig:Asymptotic_expected_limits} shows the expected and
observed limits on $\sigma\mathcal{B}(X\to\Wg)$ at 95\% confidence level (\CL)
for both the narrow- and broad-width resonance hypotheses.
The limits for the narrow-width resonance hypothesis are 15\% to 20\% better than the ones for the broad-width resonance hypothesis.

\begin{figure}[htp]
  \centering
  \includegraphics[width=0.49\textwidth]{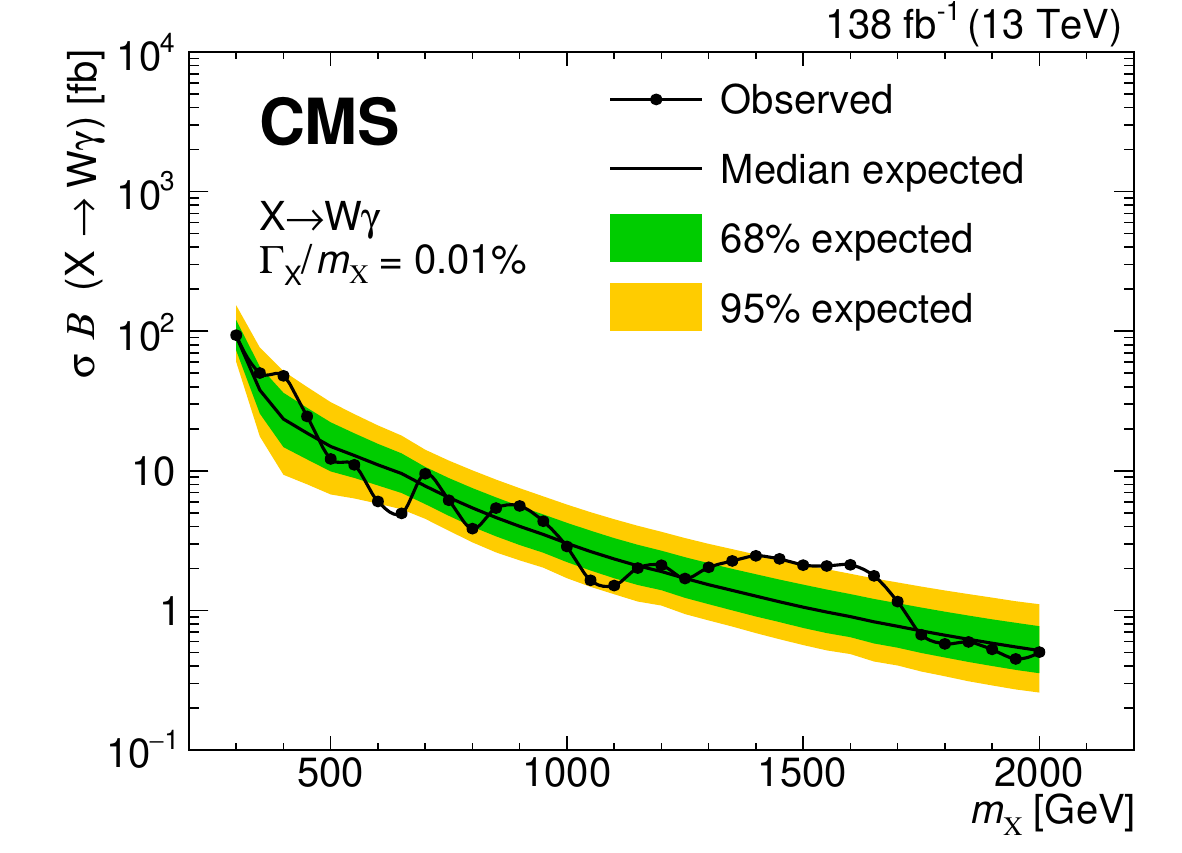}
  \includegraphics[width=0.49\textwidth]{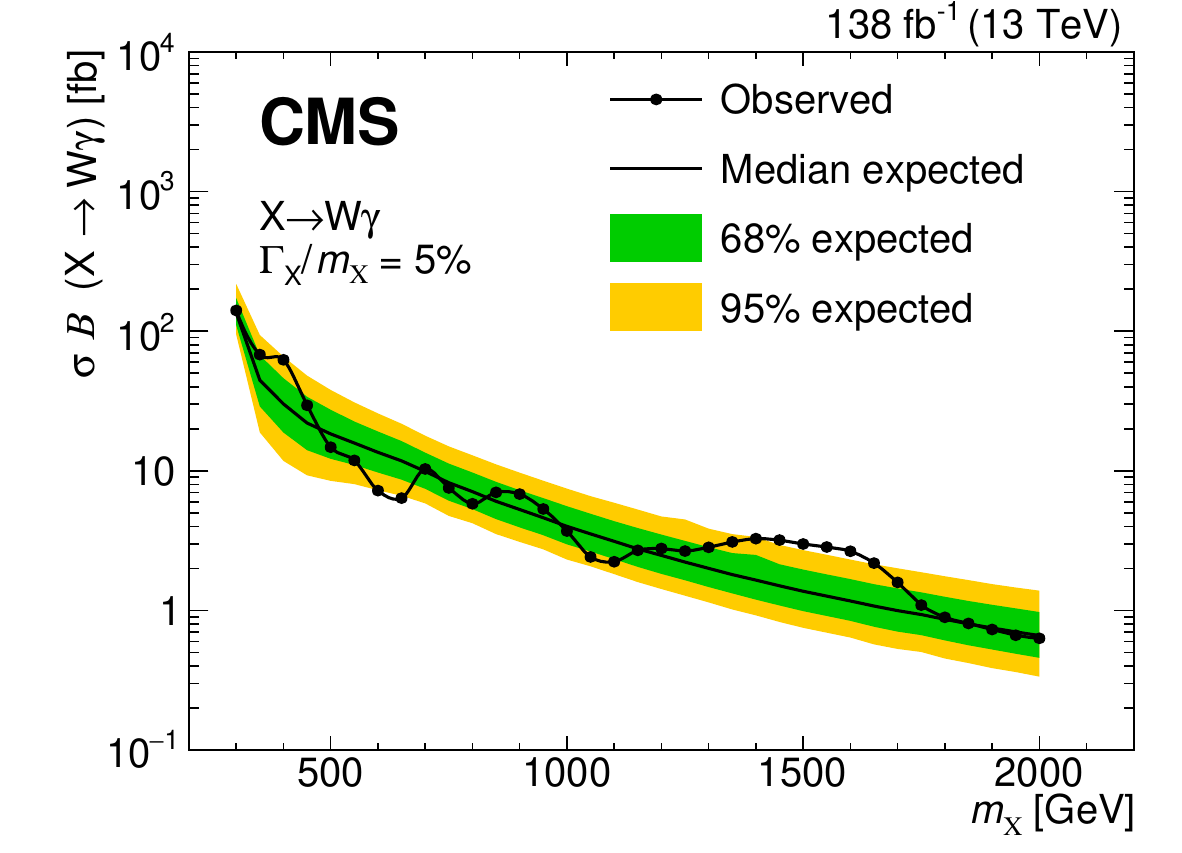}
  \caption{\label{fig:Asymptotic_expected_limits_combined} Expected and observed limits at 95\% \CL on $\sigma\mathcal{B}(X\to\Wg)$ utilizing both hadronic (Ref.~\cite{CMS:2021zxu}) and leptonic (this analysis) \PW decays as a function of the \PX resonant mass. The results for the narrow (broad)
    width assumption are shown on the left (right). 
    }
\end{figure}

\begin{figure}[htp]
  \centering
  \includegraphics[width=0.49\textwidth]{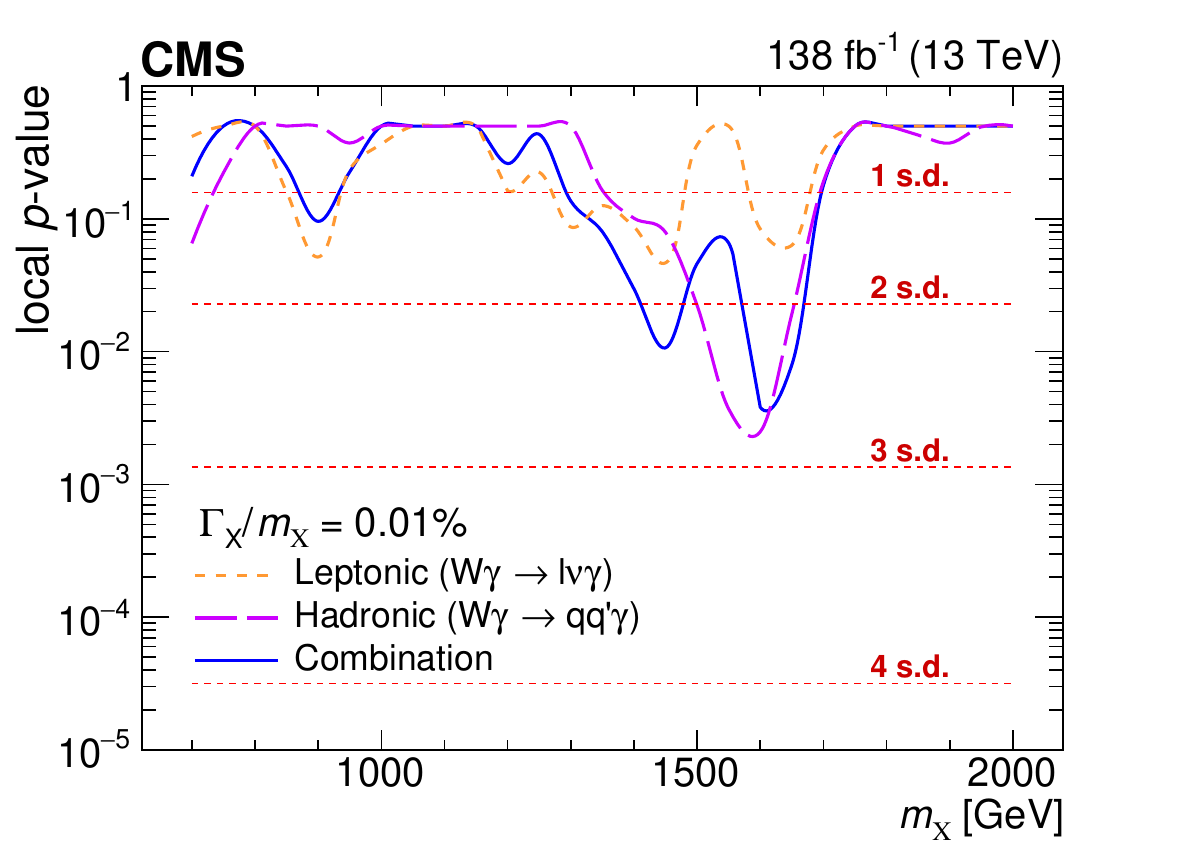}
  \includegraphics[width=0.49\textwidth]{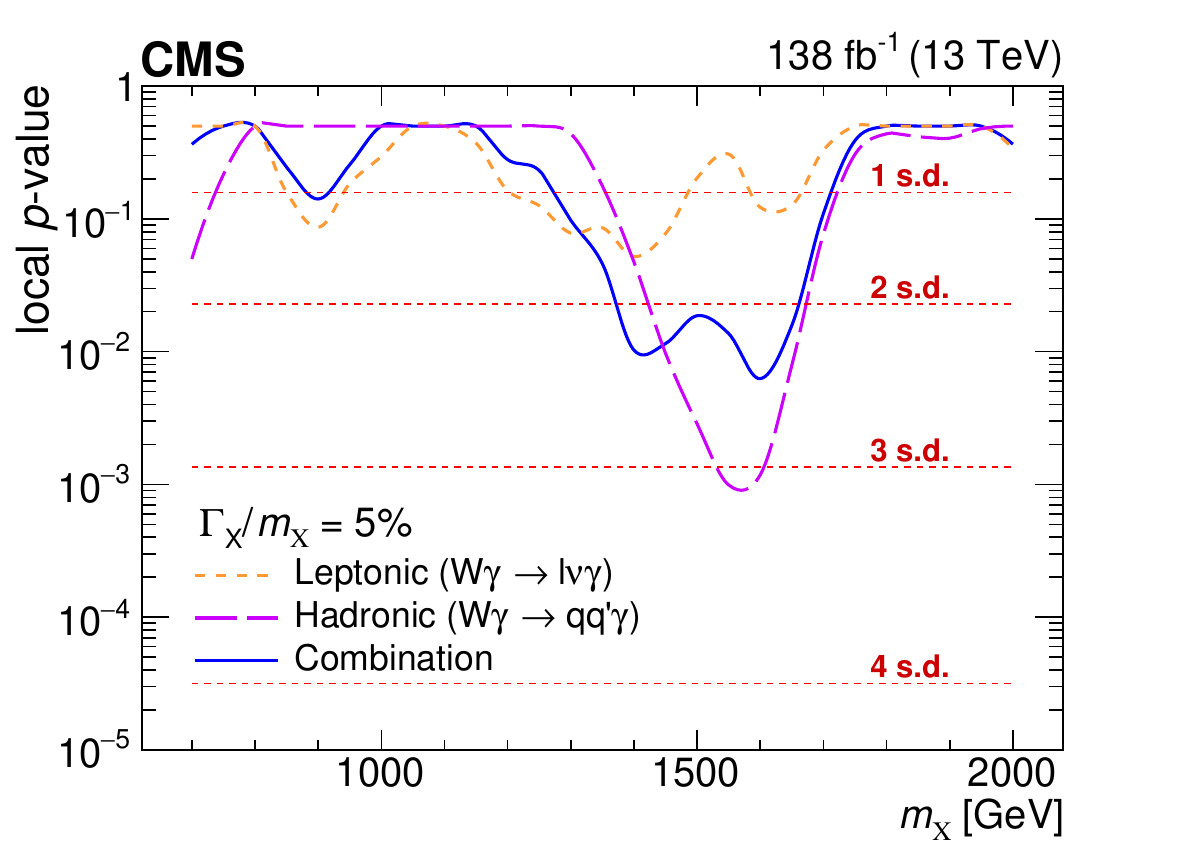}
  \caption{\label{fig:local_p_value} 
    Observed local $\it{p}$-values for narrow (left) and broad (right) resonance width hypotheses with the background-only fit in the hadronic (Ref.~\cite{CMS:2021zxu}) and leptonic (this analysis) channels. The blue line shows the observed local $\it{p}$-values after their combination. In the hadronic channel (violet line), the largest excess corresponds to a local significance of 2.8 (3.1) standard deviations (s.d.) for narrow (broad) signals. In the leptonic channel (orange line), the largest local significance is 1.7 (1.6) standard deviations. After combining with the leptonic channel, the largest excess is 2.7 (2.5) standard deviations.
    }
\end{figure}

An earlier study based on the hadronic decay
mode of the \PW boson using the same 13\TeV data
sample~\cite{CMS:2021zxu} identified a small
excess of events over background expectations near a \Wg resonance
mass of 1.58\TeV, with 2.8 (3.1) standard deviations of local
significance for the narrow (broad) signal hypotheses.
The event selections of this search and the previous one are orthogonal;
only the integrated luminosity uncertainties are correlated.
In this search, the largest local significance for the narrow (broad) signal hypotheses is 1.7 (1.6) standard deviations.
At 1.58\TeV, the two analyses agree within 1.8 standard deviations.
The combined limits on $\sigma\mathcal{B}(\PX\to\Wg)$ are shown in Fig.~\ref{fig:Asymptotic_expected_limits_combined}.
The local $\it{p}$-values for the background-only fit are presented in Fig.~\ref{fig:local_p_value}, with the left (right) plot showing the results for the narrow (broad) resonance hypothesis.
Both separate and combined $\it{p}$-values are shown over the common mass range.
The maximal excess observed in the earlier study is reduced to 2.7 (2.5) standard deviations of local significance for the narrow (broad) signal hypotheses.

\section{Summary}

This study presents a search for a new particle \PX
decaying into a \PW boson and a photon with mass hypotheses from 0.3
to 2.0\TeV and \PX width-to-mass ratio hypotheses of 0.01\% (narrow) and 5\% (broad).
Events with a muon or an electron, large \ptmiss, and a
high-\pt photon are analyzed. The transverse mass of the lepton,
photon, and \ptmiss is the primary kinematic observable. The search utilizes proton-proton collision data
collected at a center-of-mass energy of 13\TeV with the CMS detector
at the LHC throughout 2016--2018, corresponding to an integrated
luminosity of 138\fbinv.

This search reveals no statistically significant excess of events
above the background. Upper limits at the 95\%
confidence level on the product of the cross section and branching
fraction for \Wg resonances are set. These limits span a range from
94 (137) to 0.75 (0.81)\unit{fb} for the narrow (broad) resonance
hypothesis. These findings represent the most stringent
constraints to date on the existence of such resonances across the probed mass range.

This search complements an earlier study based on the hadronic decay
mode of the \PW boson using the full 13\TeV data
sample~\cite{CMS:2021zxu}. By combining searches for both leptonic and hadronic decays of \PW bosons, the largest local excess seen in the hadronic channel is reduced from 3.1 to 2.5 standard deviations for the broad signal-width hypothesis. The upper limit at the 95\% confidence level at 2\TeV is reduced from 0.75 (0.81) to 0.50 (0.63)\unit{fb} for the narrow (broad) resonance hypothesis.

\begin{acknowledgments}
We congratulate our colleagues in the CERN accelerator departments for the excellent performance of the LHC and thank the technical and administrative staffs at CERN and at other CMS institutes for their contributions to the success of the CMS effort. In addition, we gratefully acknowledge the computing centers and personnel of the Worldwide LHC Computing Grid and other centers for delivering so effectively the computing infrastructure essential to our analyses. Finally, we acknowledge the enduring support for the construction and operation of the LHC, the CMS detector, and the supporting computing infrastructure provided by the following funding agencies: SC (Armenia), BMBWF and FWF (Austria); FNRS and FWO (Belgium); CNPq, CAPES, FAPERJ, FAPERGS, and FAPESP (Brazil); MES and BNSF (Bulgaria); CERN; CAS, MoST, and NSFC (China); MINCIENCIAS (Colombia); MSES and CSF (Croatia); RIF (Cyprus); SENESCYT (Ecuador); ERC PRG, RVTT3 and MoER TK202 (Estonia); Academy of Finland, MEC, and HIP (Finland); CEA and CNRS/IN2P3 (France); SRNSF (Georgia); BMBF, DFG, and HGF (Germany); GSRI (Greece); NKFIH (Hungary); DAE and DST (India); IPM (Iran); SFI (Ireland); INFN (Italy); MSIP and NRF (Republic of Korea); MES (Latvia); LMTLT (Lithuania); MOE and UM (Malaysia); BUAP, CINVESTAV, CONACYT, LNS, SEP, and UASLP-FAI (Mexico); MOS (Montenegro); MBIE (New Zealand); PAEC (Pakistan); MES and NSC (Poland); FCT (Portugal); MESTD (Serbia); MCIN/AEI and PCTI (Spain); MOSTR (Sri Lanka); Swiss Funding Agencies (Switzerland); MST (Taipei); MHESI and NSTDA (Thailand); TUBITAK and TENMAK (Turkey); NASU (Ukraine); STFC (United Kingdom); DOE and NSF (USA).
   
 \hyphenation{Rachada-pisek} Individuals have received support from the Marie-Curie program and the European Research Council and Horizon 2020 Grant, contract Nos.\ 675440, 724704, 752730, 758316, 765710, 824093, 101115353, 101002207, and COST Action CA16108 (European Union); the Leventis Foundation; the Alfred P.\ Sloan Foundation; the Alexander von Humboldt Foundation; the Science Committee, project no. 22rl-037 (Armenia); the Belgian Federal Science Policy Office; the Fonds pour la Formation \`a la Recherche dans l'Industrie et dans l'Agriculture (FRIA-Belgium); the Agentschap voor Innovatie door Wetenschap en Technologie (IWT-Belgium); the F.R.S.-FNRS and FWO (Belgium) under the ``Excellence of Science -- EOS" -- be.h project n.\ 30820817; the Beijing Municipal Science \& Technology Commission, No. Z191100007219010 and Fundamental Research Funds for the Central Universities (China); the Ministry of Education, Youth and Sports (MEYS) of the Czech Republic; the Shota Rustaveli National Science Foundation, grant FR-22-985 (Georgia); the Deutsche Forschungsgemeinschaft (DFG), under Germany's Excellence Strategy -- EXC 2121 ``Quantum Universe" -- 390833306, and under project number 400140256 - GRK2497; the Hellenic Foundation for Research and Innovation (HFRI), Project Number 2288 (Greece); the Hungarian Academy of Sciences, the New National Excellence Program - \'UNKP, the NKFIH research grants K 131991, K 133046, K 138136, K 143460, K 143477, K 146913, K 146914, K 147048, 2020-2.2.1-ED-2021-00181, and TKP2021-NKTA-64 (Hungary); the Council of Science and Industrial Research, India; ICSC -- National Research Center for High Performance Computing, Big Data and Quantum Computing and FAIR -- Future Artificial Intelligence Research, funded by the NextGenerationEU program (Italy); the Latvian Council of Science; the Ministry of Education and Science, project no. 2022/WK/14, and the National Science Center, contracts Opus 2021/41/B/ST2/01369 and 2021/43/B/ST2/01552 (Poland); the Funda\c{c}\~ao para a Ci\^encia e a Tecnologia, grant CEECIND/01334/2018 (Portugal); the National Priorities Research Program by Qatar National Research Fund;  MCIN/AEI/10.13039/501100011033, ERDF ``a way of making Europe", and the Programa Estatal de Fomento de la Investigaci{\'o}n Cient{\'i}fica y T{\'e}cnica de Excelencia Mar\'{\i}a de Maeztu, grant MDM-2017-0765 and Programa Severo Ochoa del Principado de Asturias (Spain); the Chulalongkorn Academic into Its 2nd Century Project Advancement Project, and the National Science, Research and Innovation Fund via the Program Management Unit for Human Resources \& Institutional Development, Research and Innovation, grant B37G660013 (Thailand); the Kavli Foundation; the Nvidia Corporation; the SuperMicro Corporation; the Welch Foundation, contract C-1845; and the Weston Havens Foundation (USA).
\end{acknowledgments}
\bibliography{auto_generated}
\cleardoublepage \appendix\section{The CMS Collaboration \label{app:collab}}\begin{sloppypar}\hyphenpenalty=5000\widowpenalty=500\clubpenalty=5000
\cmsinstitute{Yerevan Physics Institute, Yerevan, Armenia}
{\tolerance=6000
A.~Hayrapetyan, A.~Tumasyan\cmsAuthorMark{1}\cmsorcid{0009-0000-0684-6742}
\par}
\cmsinstitute{Institut f\"{u}r Hochenergiephysik, Vienna, Austria}
{\tolerance=6000
W.~Adam\cmsorcid{0000-0001-9099-4341}, J.W.~Andrejkovic, T.~Bergauer\cmsorcid{0000-0002-5786-0293}, S.~Chatterjee\cmsorcid{0000-0003-2660-0349}, K.~Damanakis\cmsorcid{0000-0001-5389-2872}, M.~Dragicevic\cmsorcid{0000-0003-1967-6783}, P.S.~Hussain\cmsorcid{0000-0002-4825-5278}, M.~Jeitler\cmsAuthorMark{2}\cmsorcid{0000-0002-5141-9560}, N.~Krammer\cmsorcid{0000-0002-0548-0985}, A.~Li\cmsorcid{0000-0002-4547-116X}, D.~Liko\cmsorcid{0000-0002-3380-473X}, I.~Mikulec\cmsorcid{0000-0003-0385-2746}, J.~Schieck\cmsAuthorMark{2}\cmsorcid{0000-0002-1058-8093}, R.~Sch\"{o}fbeck\cmsorcid{0000-0002-2332-8784}, D.~Schwarz\cmsorcid{0000-0002-3821-7331}, M.~Sonawane\cmsorcid{0000-0003-0510-7010}, S.~Templ\cmsorcid{0000-0003-3137-5692}, W.~Waltenberger\cmsorcid{0000-0002-6215-7228}, C.-E.~Wulz\cmsAuthorMark{2}\cmsorcid{0000-0001-9226-5812}
% ...
\par}
\vskip\cmsinstskip
\dag:~Deceased\\
$^{1}$Also at Yerevan State University, Yerevan, Armenia\\
$^{2}$Also at TU Wien, Vienna, Austria\\
% ...
\end{sloppypar}
\end{document}